\documentclass[twocolumn,floats,floatfix,showpacs,amssymb,prd,superscriptaddress,nofootinbib]{revtex4-2}

\usepackage{amssymb,amsmath,verbatim,mathtools,needspace,enumitem,etoolbox,graphicx,physics,microtype,afterpage,xspace,tabularx,lmodern,multirow,bm}
\usepackage{dcolumn}
\usepackage{multirow}
\usepackage{float}
\usepackage{booktabs}
\usepackage{gensymb}
\usepackage{lipsum}
\usepackage[dvipsnames, usenames]{xcolor}
\definecolor{linkcolor}{rgb}{0.0,0.3,0.5}
\usepackage[unicode, colorlinks=true, linkcolor=linkcolor, citecolor=linkcolor, filecolor=linkcolor, urlcolor=linkcolor, linktocpage, breaklinks]{hyperref}
\usepackage[all]{hypcap}
\usepackage[T1]{fontenc}
\usepackage[utf8]{inputenc}
\usepackage[usenames,dvipsnames]{xcolor}
\definecolor{TurkishBlue}{HTML}{144893}
\definecolor{TurkishBlue2}{HTML}{1985B6}
\hypersetup{colorlinks=true,citecolor=TurkishBlue2,linkcolor=TurkishBlue2,urlcolor=TurkishBlue2}

\definecolor{romared}{RGB}{142,0,28}
\usepackage{aas_macros}
\usepackage{makecell}
\usepackage{soul}
\usepackage[nolist,nohyperlinks]{acronym}

\usepackage{array}

\usepackage{bm}

\usepackage [english]{babel}
\usepackage [autostyle, english = american]{csquotes}
\MakeOuterQuote{"}

\newcommand{\approptoinn}[2]{\mathrel{\vcenter{
  \offinterlineskip\halign{\hfil$##$\cr
    #1\propto\cr\noalign{\kern2pt}#1\sim\cr\noalign{\kern-2pt}}}}}

\usepackage[toc,page]{appendix}

\newcommand{\be}{\begin{align}}

\newcommand{\ee}{\end{align}}

\newcommand{\msu}{eXtreme Gravity Institute, Department of Physics,\\
Montana State University, Bozeman, Montana 59717, USA}
\newcommand{\jhu}{William H.~Miller III Department of Physics and Astronomy,\\ Johns
Hopkins University, 3400 N Charles St, Baltimore, MD 21218, USA}
\newcommand{\ligo}{LIGO Laboratory, California Institute of Technology, Pasadena, CA 91125, USA}

\newcommand{\gw}{gravitational-wave }

\newcommand{\SB}[1]{\textcolor{blue}{\textbf{SB:} #1}}
\newcommand{\NP}[1]{\textcolor{red}{\textbf{NP:} #1}}

\newcommand{\MCtwo}[1]{\textcolor{orange}{\textbf{MÇ:} #1}} 

\begin{document}

\title{A compact time-frequency representation for gravitational-wave data analysis}

\author{Noah~Pearson}
\email{noahpearson1@montana.edu}
\affiliation{\msu}

\author{Mesut Çalışkan}
\email{caliskan@jhu.edu}
\affiliation{\jhu}

\author{Sophie Bini}
\email{bini@caltech.edu}
\affiliation{\ligo}

\author{Neil J.~Cornish}
\email{ncornish@montana.edu}
\affiliation{\msu}


\begin{abstract}
Time-frequency (wavelet) domain analyses are seeing greater use for \gw data analysis due to the advantages they have in handling non-stationary noise. 
A popular choice is the Wilson-Daubechies-Meyer (WDM) wavelet transform, which uses a window function that is very compact in frequency, but more spread out in time. 
In this work, we consider an alternative window function that is built from a sum of phase-shifted Gaussians. 
This ``Gaussian'' window is more symmetric in time-frequency, and consequently more compact.
We examine the properties of this window function and the implications it has for \gw analyses.
We calculate the window's time-frequency variance product analytically and verify it numerically.
For the symmetric case, where the window has the same form in time and frequency, the construction comes within $\sim2.4\%$ of saturating the Heisenberg-Gabor uncertainty limit. 
We conjecture that this Gaussian window achieves the minimum time-frequency area of any WD wavelet window.
\end{abstract}

\maketitle

\section{Introduction}

Standard \gw data analysis to-date has relied upon the assumption that the underlying noise is stationary, i.e., that it does not change over the signal duration. 
This is a good approximation for short-lived signals in today's ground-based detectors, but will fail for future detectors like the space-based Laser Interferometer Space Antenna (LISA), where massive black hole merger and galactic binary signals will be in the sensitive band for months to years~\cite{LISAredbook2024}.
To account for non-stationarity, methods have been developed to move \gw data analysis from the frequency domain into the time-frequency (wavelet) domain where non-stationary noise can be robustly incorporated into standard Bayesian inference~\cite{Necula2012,Cornish2020}.
The preferred class of wavelets are the Wilson-Daubechies (WD) family for their fast FFT-like transforms, orthogonality, and approximately diagonal noise covariance matrices for locally-stationary noise~\cite{Cornish2020,Cornish:2025awt}. 

\begin{figure}[h]
    \centering
    \includegraphics[width=\linewidth]{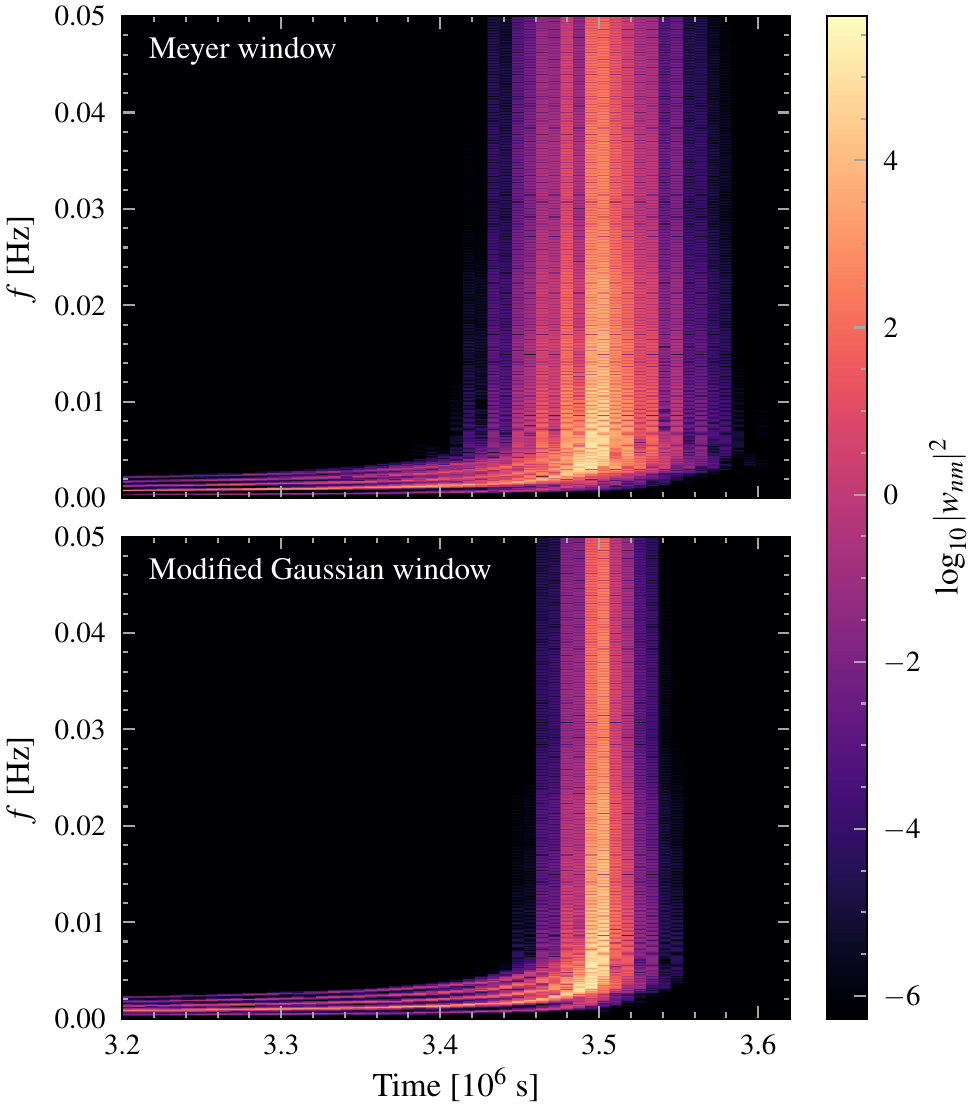}
    \caption{WD transforms of a LISA-band massive black hole binary merger, whitened by the LISA sensitivity curve so that the pixel powers $|w_{nm}|^2$ sum to the squared signal-to-noise ratio ($\mathrm{SNR} \simeq 4300$).
    The whitening uses the sky-averaged two-channel LISA sensitivity of Ref.~\cite{Robson:2018ifk}, including the galactic confusion noise for a one-year observation time.
    The panels zoom into the merger, from $2.9\times10^{5}$~s before the peak pixel column to $1.3\times10^{5}$~s after it.
    The modified Gaussian window (lower panel) produces a sharper transform than the Meyer window (upper panel), and captures $99\%$ of the signal power in about half as many pixels ($979$ against $1793$, counted over $t \ge 5\times10^{5}$~s).
    The Meyer window spreads the merger power over more pixel columns on either side of the peak than the modified Gaussian window does.}
    \label{fig:lisa_wd_transform}
\end{figure}

The Meyer window has been the default choice in the construction of a WD wavelet basis, primarily for its frequency localization.
However, by virtue of the Heisenberg-Gabor uncertainty~\cite{gabor1946}, this requires that the window's time extent is relatively large.
This gives the Wilson-Daubechies-Meyer (WDM) basis a lengthy time domain kernel even after truncation~\cite{Cornish2020}. 
While not inherently problematic, this means that, e.g., the impact of gaps or ephemeral noise corruption in data extends over many wavelets in time~\cite{Pearson2026}.
However, the window is the primary freedom in the WD basis construction.
Any window satisfying the orthogonality condition of App.~\ref{appx: POU and orthog} yields a lossless transform with the same fast implementation and the same noise-covariance properties~\cite{Cornish2020,Cornish:2025awt,Johnson2026}, so the window can be optimized for time-frequency localization without affecting either.
The optimal choice has remained an open question~\cite{Cornish:2025awt}.
Since the corruption caused by a gap or a glitch extends over the full time span of the window kernel~\cite{Pearson2026}, a window that is more compact in time directly shrinks the corrupted region of the time-frequency plane.
Here, we investigate a `modified' Gaussian window function~\cite{Daubechies1991}, a lattice sum of phase-shifted Gaussians orthogonalized so that the resulting WD wavelet basis is orthonormal (a single true Gaussian cannot generate one).
For brevity, we call this window the ``Gaussian'' window and its associated transform, ``WDG.''
Compared to the Meyer window, the resulting construction has a modest broadening in frequency 
and nearly twice the compactness in time. 
This time-frequency compactness may be particularly useful for low-latency analysis of black hole mergers or data with gaps and glitches.

In this article, we correct the plots of the Gaussian window function shown in the original paper~\cite{Daubechies1991} (an error also noticed by Floris \& de Hon~\cite{floris2018}), clarify the notation for the window definition, and provide code to implement the transformation.\footnote{\url{https://github.com/XGI-MSU}}
We compare the compactness of the Gaussian (WDG) transform with the Meyer (WDM) transform for different choices of the window parameters.
Because of the similarity between the discrete WDG transform and the continuous Morlet-Gabor transform, which saturates the Heisenberg-Gabor uncertainty limit, we conjecture that the WDG transform is the most compact choice in the WD family, and quite possibly, for any discrete wavelet transform.
In support of this conjecture we show that the time-frequency symmetric Gaussian window comes within $\sim$2.4\% of saturating the Heisenberg-Gabor uncertainty limit.
We further argue that the structure of the analytic result, rather than the size of the excess alone, suggests that little room is left for a more compact window within the WD family.
Since no window can go below the bound, no WD window can improve on this product by more than $2.4\%$.

Figure~\ref{fig:lisa_wd_transform} shows the WD-transforms of a simulated massive black hole binary observed by LISA.
The system has detector frame masses $m_1 = 2\times 10^5\; M_\odot$ and $m_2 = 1\times 10^5\; M_\odot$ and dimensionless spins $\chi_1= 0.42$ and $\chi_2=0.85$. 
The waveforms are whitened using the LISA sensitivity curve from Ref.~\cite{Robson:2018ifk}.
To capture 99\% of the signal-to-noise squared the Meyer transform uses 1793 pixels, while the Gaussian window uses just 979 pixels. 
The difference is particularly large around merger, where the Gaussian transform is far more compact.

This paper is organized as follows:
in Sec.~\ref{sec: definition}, we define the Gaussian window, give the recursion for its coefficients, and discuss the shape parameter $\nu$ that trades localization in time against localization in frequency.
In Sec.~\ref{sec:uncertainty}, we examine the window's compactness, computing the time-frequency variance product analytically and numerically, comparing it with the Meyer window and with the Heisenberg-Gabor limit.
In Sec.~\ref{sec:implications}, we discuss the implications for \gw analyses, in particular the localization of transient signals and the reduced spread of spectral leakage from data gaps and edges.
There, we also give the leading terms of the noise covariance matrix in the WDG basis.
In App.~\ref{appx: POU and orthog}, we verify numerically that the window satisfies orthogonality and partition of unity for the standard truncations of the construction. 
In App.~\ref{appx: PSF}, we show the point spread function for both the Gaussian and Meyer windows, as an alternative lens for compactness.

\section{The WDG wavelet transform}
\label{sec: definition}

Daubechies \textit{et al.} originally showed that WD window functions with exponential decay in both time and frequency can be built using the Gaussian kernel~\cite{Daubechies1991},
\[
  g^\nu(f)=(2\nu)^{1/4}e^{-\nu\pi f^2},
\]
Defining,

\begin{equation}
  g_{mn}(f) = e^{2\pi i \alpha m f}\, g(f-\beta n),
  \qquad m,n\in\mathbb{Z},
\end{equation}

\noindent where $\alpha = 1/2$ and $\beta = 1$.
The window construction is,

\begin{equation}
  \tilde\phi(f)=\frac{2}{\sqrt{A_\nu+B_\nu}}
  \sum_{m,n\in\mathbb{Z}} a_{mn}\,g_{mn}^\nu(f).
  \label{eq:gaussian window}
\end{equation}

\noindent We use the above definition to clarify the inconsistent frequency domain notation used in Ref.~\cite{Daubechies1991}.
The coefficients $a_{mn}$ are defined recursively by,

\begin{equation}
\begin{aligned}
  a_{mn}
  &= \sum_{k=0}^{\infty}\frac{(2k)!}{2^{2k}(k!)^2}\,b_{mn}^k, \\
  b_{mn}^k
  &= b_{mn}^{k-1}
     -\frac{2}{A_\nu+B_\nu}
       \sum_{(m',n')}
       \omega_{mn,m'n'}\,b_{m'n'}^{k-1}, \\
  \omega_{mn,m'n'}
  &= \exp\left[
      i(m'-m)(n+n')\frac{\pi}{2} \right.\\
   &   \left. 
   \quad \quad -\frac{\nu\pi}{2}(n-n')^2
      -\frac{\pi}{8\nu}(m-m')^2
     \right], \\
  b_{mn}^{0} &= \delta_{m0}\delta_{n0}.
  \label{eq: coeffs 1}
\end{aligned}
\end{equation}

\noindent The lattice constants for the kernel $\omega_{mn,m'n'}$ above are given by the matrix of inner products of $g^{\nu}_{mn}$.
Here,
\begin{equation}
\begin{aligned}    
&A_\nu  =
\inf_{a,b \in [0,1]}
\left[
\left|(U_Z g^\nu)(a,b)\right|^2
+
\left|(U_Z g^\nu)\left(a,b+\frac{1}{2}\right)\right|^2
\right],\\
&B_\nu = 
\sup_{a,b \in [0,1]}
\left[
\left|(U_Z g^\nu)(a,b)\right|^2
+
\left|(U_Z g^\nu)\left(a,b+\frac{1}{2}\right)\right|^2
\right],\\
\end{aligned}
\end{equation}
\noindent where $(U_Z g)$ refers to the Zak transform, which is defined as~\cite{Daubechies1991},
\begin{equation}
  (U_Z g)(a,b)
  = \sqrt{2}\sum_{k\in\mathbb{Z}} e^{2\pi i k a}\,g\bigl(2(b-k)\bigr).
\end{equation}

\begin{figure*}
    \centering
    \includegraphics[width=\linewidth]{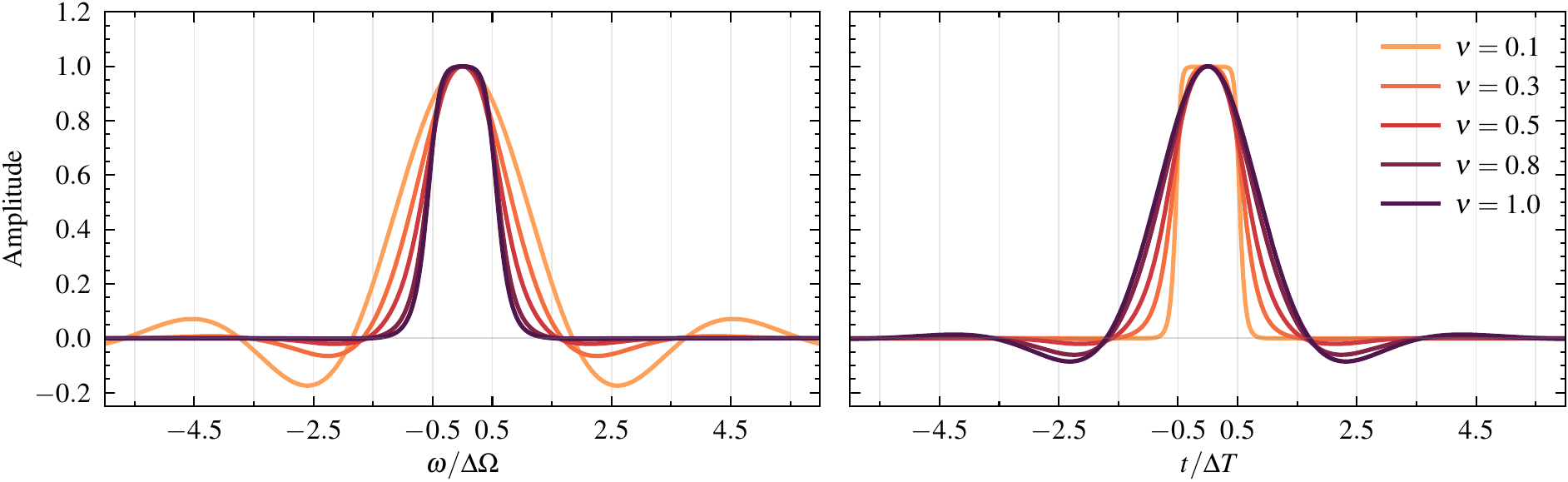}
    \caption{The modified Gaussian window in frequency (\textit{left}) and time (\textit{right}) domain varying the parameter $\nu$.
    Vertical gray lines mark bin edges.
    A low $\nu$ corresponds to a window compact in time and with a broad extent in frequency.
    A high $\nu$ corresponds to a window compact in frequency and extended in time domain.
    In the rest of this work, we will use $\nu=0.5$.}
    \label{fig:wilson_window}
\end{figure*}

Because the reconstructed window converges at least as fast as a geometric series, only a few terms in the sums are needed for high accuracy~\cite{Daubechies1991}.
We find $\mathcal{O}(20)$ to be sufficient for our use, see App.~\ref{appx: POU and orthog} and Fig.~\ref{fig:POU}.
In implementing the sums numerically, the indices are taken to range over truncated indices $n\in[-N,N]$, $m\in[-M,M]$, and $k\in[0,K_\mathrm{max}]$.
These truncations are set for $\nu = 1/2$, where the series of Eq.~\eqref{eq: coeffs 1} converges quickly, while $\nu$ far from $1/2$ needs more terms (Sec.~\ref{sec:numerical_variance}).

The WD family of wavelet transforms can be represented using a 2D grid of pixels whose values represent the wavelet amplitudes $w_{nm}$ or energy  $|w_{nm}|^2$.
The size of these pixels in time $\Delta T$ and frequency $\Delta F$ ($\Delta \Omega =2\pi\Delta F$) are constrained by $\Delta T\Delta F = 1/2$.
In the above definitions, all frequencies are in units of the wavelet pixel frequency width, $f = f/\Delta F$, as well as $\alpha$ and $\beta$.
See Ref.~\cite{Cornish2020} for more details of this representation.

Figure~\ref{fig:wilson_window} shows the  Gaussian window in the frequency and time domains.
This is a correction to Ref.~\cite{Daubechies1991}, where the plots shown of this window are inaccurate (as also found by Floris \& de Hon~\cite{floris2018}).
For low $\nu$ ($\nu=0.1$), the Gaussian window is compact in the time domain, with a flat top spanning about one $\Delta T$ pixel and the tails one adjacent pixel, while it extends for several pixels in the frequency domain.
Increasing $\nu$, the window broadens in the time domain, and gets more compact in the frequency domain.
For $\nu=0.5$, the Gaussian window has the same shape and extension in both the time and frequency domains.
Because of this symmetry, we choose $\nu=0.5$ as the standard choice used throughout this work. 

\section{Uncertainty Relation}
\label{sec:uncertainty}

Gabor in 1946 discussed the uncertainty relation for a function’s variance in time and frequency, hearkening it to the better-known Heisenberg uncertainty relation in quantum mechanics~\cite{gabor1946}.
He found that the Gaussian distribution is the solution that minimizes the time-frequency variance product (as do coherent states in quantum mechanics). 
While the ``Gabor atom'' has found its use in \gw astronomy~\cite{Cornish2014,Mathur2025}, it is not orthogonal and therefore not a unitary transformation (which will result in information loss with repeated transformations).
Gabor windows can be constructed to asymptotically approximate a lossless transform, but this becomes computationally intensive depending on the order of desired accuracy; we would prefer to work with orthogonal transforms. 

The time-frequency variance product is a standard measure for time-frequency localization~\cite{Sharma2017} and is bounded from below by the Heisenberg-Gabor uncertainty~\cite{gabor1946},

\begin{equation}
    \sigma_t\sigma_f \geq \frac{1}{4\pi}\,,
\end{equation}

\noindent where we define the variance using the window energy densities $|\tilde{\phi}(f)|^2$,

\begin{equation}
    \begin{aligned}
         & \sigma_t^2 := \mathrm{Var}[\phi(t)] = \int t^2|\phi(t)|^2\, \dd t\\
        & \sigma_f^2 := \mathrm{Var}[\tilde{\phi}(f)] = \int f^2|\tilde{\phi}(f)|^2\, \dd f\\\
    \end{aligned}
    \label{eq:variances}
\end{equation}

It is desirable for many wavelet analyses to work in a basis that is compact in time-frequency (a smaller variance product), so as to not spread energy too far from its point source origin in time-frequency.
Orthogonal wavelet filters have been designed parametrically to optimize the variance product to arbitrary precision~\cite{Sharma2017}.
In this work, we conjecture that the Gaussian window achieves the minimum time-frequency area of any WD wavelet window, motivated by how closely it approaches the Heisenberg-Gabor bound.
We first evaluate the variance product of the Gaussian window analytically (Sec.~\ref{sec:analytic_variance}), and then verify the result numerically and compare it with the Meyer window (Sec.~\ref{sec:numerical_variance}).

\subsection{Analytic Variance Product}
\label{sec:analytic_variance}

The variance product of the Gaussian window can be evaluated analytically, since every moment of the window is a Gaussian integral.
Two observations reduce the calculation to elementary steps.
First, no separate normalization of Eq.~\eqref{eq:variances} is needed, because integrating the $j=0$ case of Eq.~\eqref{eq:DJJcondition} over a single frequency pixel gives $\int |\tilde{\phi}(f)|^2\, \dd f = 1$ exactly.
Second, the product of any two of the lattice Gaussians in Eq.~\eqref{eq:gaussian window} is again a Gaussian, so every cross term of the second moment is elementary,
\begin{equation}
    \begin{aligned}
    \int f^2\, g^{\nu\,*}_{mn}(f)\, g^{\nu}_{m'n'}(f)\, \dd f
    &= \omega_{mn,m'n'} \\
    &\quad\times\left[\left(c+i\kappa\right)^{2}+\frac{1}{4\pi\nu}\right],
    \end{aligned}
    \label{eq:moment_element}
\end{equation}
with $c=(n+n')/2$, $\kappa=(m'-m)/(4\nu)$, frequencies in units of $\Delta F$, and $\omega_{mn,m'n'}$ the overlap kernel of Eq.~\eqref{eq: coeffs 1}, which is precisely the matrix of inner products of the $g^{\nu}_{mn}$.
The bracket is the second moment of the product Gaussian, the square of its center $c + i\kappa$ (the relative modulation shifts the center into the complex plane) plus its variance $1/(4\pi\nu)$.
The window is real and even, so its first moment vanishes and Eq.~\eqref{eq:variances} is the central variance.
Summing Eq.~\eqref{eq:moment_element} against the window coefficients and dividing by the norm gives the frequency variance in closed form,
\begin{widetext}
\begin{equation}
    \frac{\sigma_f^2}{\Delta F^2}
    = \frac{1}{4\pi\nu}
    + \frac{\displaystyle\sum_{mn,\,m'n'} a^{*}_{mn}\, a_{m'n'}\,\omega_{mn,m'n'}\left[\frac{n+n'}{2}+\frac{i\,(m'-m)}{4\nu}\right]^{2}}
           {\displaystyle\sum_{mn,\,m'n'} a^{*}_{mn}\, a_{m'n'}\,\omega_{mn,m'n'}}\,,
    \label{eq:sigmaf_analytic}
\end{equation}
\end{widetext}

\noindent the variance of a single Gaussian plus a lattice correction from the orthogonalization, with both sums running over the truncated index lattice of Eq.~\eqref{eq:gaussian window}.
The correction is a ratio of rapidly converging Gaussian sums, determined by the truncations of App.~\ref{appx: POU and orthog} to machine precision at the standard choice $\nu = 1/2$ (the coefficient series converges more slowly at small $\nu$, see Sec.~\ref{sec:numerical_variance}).

The time domain, on the other hand, requires no separate computation.
Taking the Fourier transform of the window term by term exchanges the roles of the translations and the modulations, and inverts the width parameter of the Gaussian kernel.
Relabeling the lattice indices shows that the time-domain window at shape parameter $\nu$ is an exact rescaled copy of the frequency-domain window at the dual parameter $1/(4\nu)$,
\begin{equation}
    \phi_{\nu}(t) = \sqrt{2}\,\tilde\phi_{1/(4\nu)}\!\left(2\,t\,\Delta F\right).
    \label{eq:nu_duality}
\end{equation}
This rescaling identity was noted in Ref.~\cite{Daubechies1991}, and we re-derive and verify it under our conventions.
Under the relabeling $(m, n) \rightarrow (n, -m)$ the kernel $\omega_{mn,m'n'}$ of Eq.~\eqref{eq: coeffs 1} maps onto itself up to the sign $(-1)^{mn + m'n'}$, and $A_\nu + B_\nu$ is invariant, so the coefficients obey $a_{n,-m}(1/(4\nu)) = (-1)^{mn}\, a_{mn}(\nu)$ and Eq.~\eqref{eq:nu_duality} holds at any $M = N$.
The time variance follows from Eq.~\eqref{eq:sigmaf_analytic} by the substitution $\nu \rightarrow 1/(4\nu)$, since setting $u = 2 t \Delta F$ in Eq.~\eqref{eq:variances}, together with $\Delta T \Delta F = 1/2$, converts the rescaling into pixel units.
Hence, we have
\begin{equation}
    \frac{\sigma_t}{\Delta T}\bigg|_{\nu} = \frac{\sigma_f}{\Delta F}\bigg|_{1/(4\nu)}\,.
    \label{eq:sigma_swap}
\end{equation}
Equation~\eqref{eq:nu_duality} displays the role of the shape parameter explicitly.
Away from the symmetric choice, the window improves localization in one domain at the cost of the other, as expected.
For instance, at $\nu=0.3$, we find $\sigma_f = 0.52\,\Delta F$ against $\sigma_t = 0.34\,\Delta T$, consistent with the trend of Fig.~\ref{fig:wilson_window}.

It is vital to note that the variance product itself is invariant under $\nu \rightarrow 1/(4\nu)$ by Eq.~\eqref{eq:sigma_swap}, so it depends on $\nu$ only through $|\ln(2\nu)|$, and it is stationary at the choice of $\nu=1/2$ (which is self-dual), the minimum seen in Fig.~\ref{fig:area}.
Equation~\eqref{eq:nu_duality} also shows that partition of unity in time at $\nu$ is equivalent to partition of unity in frequency at the dual parameter.
At $\nu=1/2$ the window has equal widths $\sigma_f = 0.404\,\Delta F$ and $\sigma_t = 0.404\,\Delta T$, and we find
\begin{equation}
    \sigma_t \sigma_f \big|_{\nu=1/2} = 1.0236\times\frac{1}{4\pi}\,,
    \label{eq:analytic_product}
\end{equation}
$2.36\%$ above the Heisenberg-Gabor bound, the minimum over the range $0.1 \leq \nu \leq 1.5$ that we scan (Fig.~\ref{fig:area}).
Since the two widths are equal at this point, the product is $1/(4\pi)$ plus half the lattice correction, so the excess over the bound comes entirely from the correction that orthogonality demands.
The excess is strictly positive, because only a pure Gaussian saturates the bound~\cite{gabor1946} and no single Gaussian can satisfy the orthogonality condition of Eq.~\eqref{eq:DJJcondition}, which for $j \neq 0$ requires a sum of products of window translates to vanish (impossible for a window that is nowhere zero and of a single sign).

\subsection{Numerical Results and Comparison with the Meyer Window}
\label{sec:numerical_variance}

\begin{figure*}
    \centering
    \includegraphics[width=\linewidth]{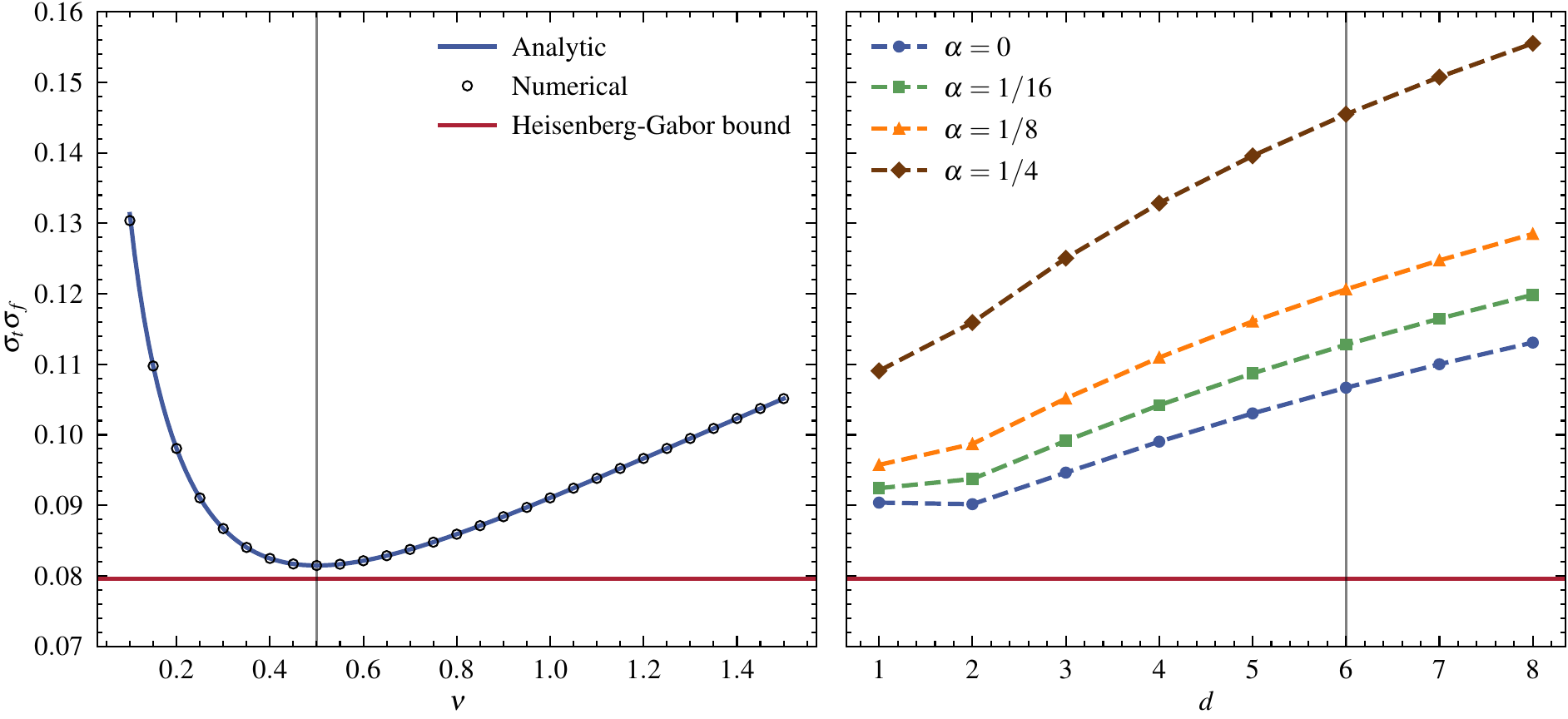}
    \caption{The time-frequency area $\sigma_t \sigma_f$ (the product of the root-mean-square widths of Eq.~\eqref{eq:variances}) of the modified Gaussian window (\textit{left}) and of the Meyer window (\textit{right}), on a shared vertical axis.
    \textit{Left:} the blue solid curve is the analytic product of Eq.~\eqref{eq:sigmaf_analytic} against the shape parameter $\nu$, and the black open circles its numerical evaluation with the standard truncations of Sec.~\ref{sec: definition}.
    \textit{Right:} the Meyer product against the steepness parameter $d$ for the flat-top fractions $\alpha = A/\Delta\Omega = 0$ (blue circles), $1/16$ (green squares), $1/8$ (orange triangles), and $1/4$ (brown diamonds), joined by dashed lines, with $2A + B = \Delta\Omega$.
    In both panels the red horizontal line marks the Heisenberg-Gabor bound $1/(4\pi)$ and the gray vertical line the standard choice, $\nu = 0.5$ and $d = 6$.
    The modified Gaussian product is minimal at $\nu = 1/2$, $2.4\%$ above the bound, while the Meyer product is $0.107$ at its standard choice ($d = 6$, $\alpha = 0$), about $34\%$ above the bound, and grows with $\alpha$ and, for $d \geq 2$, with $d$.
    }
    \label{fig:area}
\end{figure*}

We verify Eq.~\eqref{eq:sigmaf_analytic} numerically by sampling the window of Sec.~\ref{sec: definition}, with its standard truncations, on the frequency grid of a $256 \times 256$ pixel transform, transforming it to the time domain with an inverse FFT, and integrating Eq.~\eqref{eq:variances} on the two grids.
We keep the window out to $16$ pixel widths in each domain, beyond which the integrals have converged.
The left panel of Fig.~\ref{fig:area} shows the analytic product as a function of $\nu$ together with the numerical values, which agree to better than $10^{-5}$ for $\nu \geq 0.15$.
At $\nu = 0.1$ the numerical value falls $0.7\%$ short, which is due not to an integration error but to the truncation of the coefficient series of Eq.~\eqref{eq: coeffs 1} at $K_{\max} = 40$ terms.
The series of Eq.~\eqref{eq: coeffs 1} converges geometrically in the window, with ratio $(B_\nu - A_\nu)/(B_\nu + A_\nu)$ per term, which is $0.17$ at $\nu = 1/2$ but $0.92$ at $\nu = 0.1$.
Evaluating Eq.~\eqref{eq:sigmaf_analytic} with the same truncated coefficient series reproduces the deficit.
The minimum lies at the symmetric choice $\nu = 1/2$, $2.4\%$ above the bound (Eq.~\eqref{eq:analytic_product}), and the product rises on both sides as the window becomes more compact in one domain and broader in the other.

\begin{figure*}
    \centering
    \includegraphics[width=\linewidth]{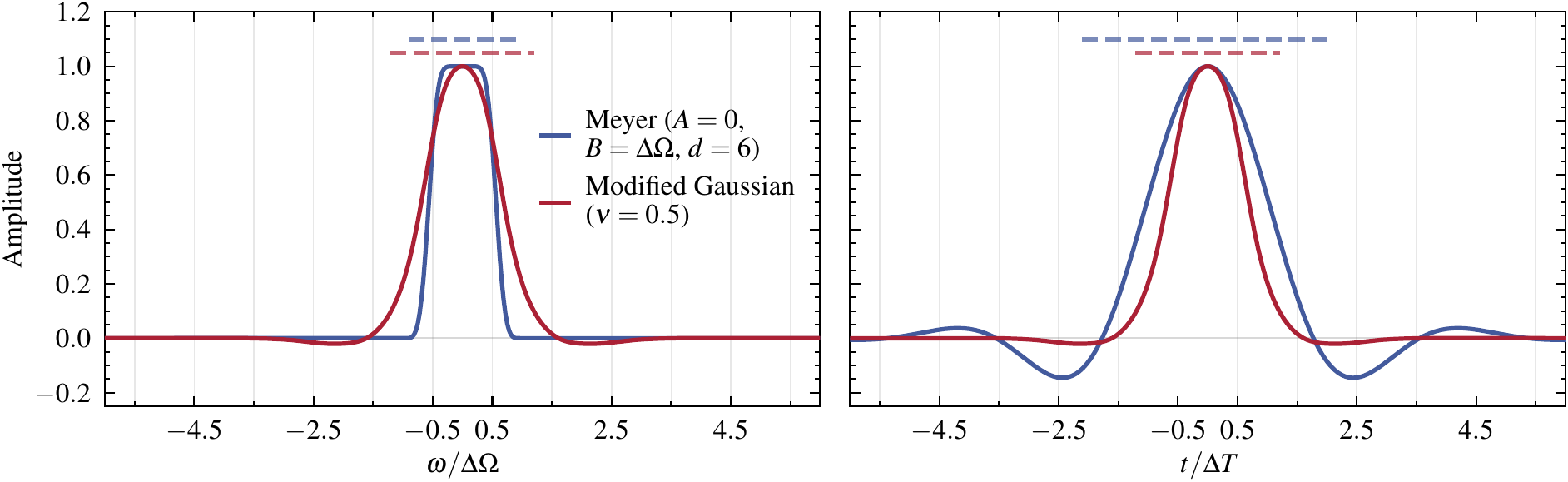}
    \caption{The modified Gaussian window with $\nu = 0.5$ (red solid) and the Meyer window with $A = 0$, $B = \Delta\Omega$, and $d = 6$ (blue solid) in the frequency (\textit{left}) and time (\textit{right}) domains, each normalized to unit peak, against frequency and time in units of the pixel widths $\Delta\Omega$ and $\Delta T$.
    The gray vertical lines mark the pixel edges.
    The red (blue) dashed bars above the curves span $\pm 3\sigma$ for the modified Gaussian (Meyer) window, with $\sigma$ the root-mean-square width of Eq.~\eqref{eq:variances}, that is $\pm 1.21$ pixel widths for the modified Gaussian window in both domains, and $\pm 0.91$ in frequency and $\pm 2.11$ in time for the Meyer window.
    The modified Gaussian window is more compact in time than the Meyer window and broader in frequency.
    }
    \label{fig:wilson_vs_gaussian}
\end{figure*}

Figure~\ref{fig:wilson_vs_gaussian} compares the Meyer with the Gaussian window in the time and frequency domains. 
For the Meyer window, we consider the shape parameters $A=0$, $B =\Delta \Omega$, and $d=6$.
The parameters $A,\,B$ are matched such that $2A+B=\Delta \Omega$, and determine the flatness and width of the window, and $d$ the sharpness of the rolloff; see Refs.~\cite{Cornish2020,Johnson2026} for more detail. 
For the Gaussian window, $\nu=0.5$. 
We measure the extent of each window by the root-mean-square widths of Eq.~\eqref{eq:variances}, the same quantities that enter the area.

In the time domain, the Gaussian window is more compact than the Meyer window ($\sigma_t = 0.40\,\Delta T$ compared to $0.70\,\Delta T$), while in the frequency domain it is broader ($\sigma_f = 0.40\,\Delta F$ compared to $0.30\,\Delta F$).
The dashed horizontal bars above the curves in Fig.~\ref{fig:wilson_vs_gaussian} span $\pm 3\sigma$ in each domain ($\pm 1.21$ pixel widths for the Gaussian window in both domains, and $\pm 0.91\,\Delta F$ and $\pm 2.11\,\Delta T$ for the Meyer window).
In practice, $99\%$ of the energy of the modified Gaussian window lies within $\pm 1.01$ pixel widths of its center in both domains, so a wavelet occupies its central pixel and the two adjacent ones.
The Meyer window reaches $99\%$ within $\pm 0.6\,\Delta F$ (its frequency support ends at $\pm \Delta F$), but only within $\pm 2.6\,\Delta T$, and $0.3\%$ of its energy falls outside $\pm 3\,\Delta T$.
By the rms measure the time extent of the Gaussian window is therefore shorter by a factor of $1.74$, at the cost of broadening by a factor $1.33$ in frequency.
These extents characterize the wavelet itself and are distinct from the pixel counts of the point spread function in App.~\ref{appx: PSF}.

The right panel of Fig.~\ref{fig:area} shows the product for the Meyer window as a function of the steepness parameter $d$, for four values of the flat-top fraction $\alpha = A/\Delta\Omega$ with $2A + B = \Delta\Omega$, so that $\alpha = 0$ is the standard choice above and $\alpha \rightarrow 1/2$ the sharp box.
The Meyer window has compact support in frequency, so we evaluate its variances by direct quadrature.

The time variance is obtained from the derivative of the frequency-domain window through Parseval's theorem, which avoids truncating its algebraic time tails.
The product grows with $\alpha$ at every $d$, and with $d$ for $d \geq 2$, because a steeper roll-off in frequency spreads the window in time (for $\alpha = 0$ the time variance grows from $0.30\,\Delta T^2$ at $d = 2$ to $0.57\,\Delta T^2$ at $d = 8$).
The exception at $d = 1$ for $\alpha = 0$ arises because the gentlest roll-off gives that window the largest frequency variance of the family ($0.131\,\Delta F^2$ compared to $0.108\,\Delta F^2$ at $d = 2$), which outweighs its smallest time variance ($0.25\,\Delta T^2$ compared to $0.30\,\Delta T^2$), so the product at $d = 1$ lies $0.2\%$ above the $d = 2$ value.
\emph{Every Meyer choice lies above the modified Gaussian minimum.}
At the standard choice, $d = 6$ and $\alpha = 0$, we find $\sigma_t \sigma_f = 0.107$, about $34\%$ above the bound.
The smallest value, at $d = 2$ and $\alpha = 0$, is $0.090$, $13\%$ above the bound.

\begin{figure*}
    \centering
    \includegraphics[width=\linewidth]{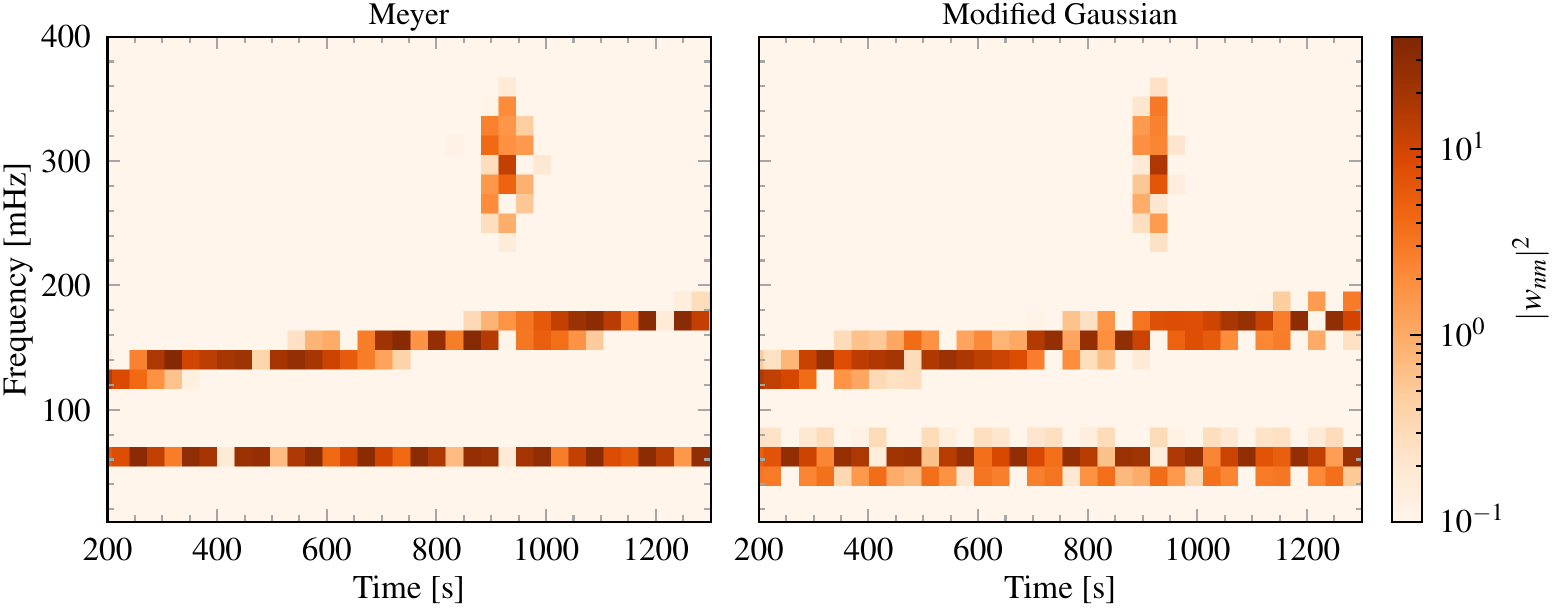}
    \caption{Time-frequency representation computed with the Meyer window (\textit{left}) and the modified Gaussian window (\textit{right}) for different kinds of signals (monochromatic signal, a chirp signal and short-duration burst).
    The modified Gaussian window is more compact in time, providing a better time localization of short-duration signal, but it is broader in frequency domain, providing a less compact representation of monochromatic and chirp signals.
    }
    \label{fig:WDM_WDG}
\end{figure*}

\section{Implications for Analysis}
\label{sec:implications}

The compactness of the Gaussian window has two direct consequences for WD-based analyses (compared to the Meyer window).
First, energy is better concentrated in time-frequency.
In Fig.~\ref{fig:lisa_wd_transform}, the same $99\%$ of the signal power fits in about half as many pixels, with the largest difference around the merger where the frequency evolves rapidly.
This is useful for low-latency analyses that act on a small set of bright pixels.
Second, the window kernels are shorter in time.
The Gaussian window decays exponentially in both domains~\cite{Daubechies1991}, while the frequency-compact Meyer window decays only algebraically in time~\cite{Johnson2026}, so a truncated time-domain kernel of a given accuracy is shorter for the Gaussian window.
In Fig.~\ref{fig:lisa_wd_transform} this appears as the power spread beyond three pixel columns after the merger within the frame ($9\times10^{-4}$ of the total for the Meyer window compared to $7\times10^{-7}$ for the Gaussian window).
This in turn shrinks everything that scales with the kernel length $K$, in particular the span of data corrupted by a gap or a glitch~\cite{Pearson2026} and the cost of a direct time-domain transform which scales like $\sim K \mathrm{log}(K)$~\cite{Cornish2020}.
A time-domain kernel that retains $99\%$ of the window energy spans $\pm 1.01\,\Delta T$ for the Gaussian window and $\pm 2.6\,\Delta T$ for the Meyer window (a factor of $2.6$), and the factor is about $3$ at $99.9\%$.
At both accuracies the saving exceeds the ratio of rms widths of $1.74$, because the Meyer window holds more of its energy beyond a few rms widths than the Gaussian window does (its $99\%$ half-width is $3.7$ rms widths, against $2.5$ for the Gaussian window).

These advantages come at the cost of localization in frequency, where the window support extends over more frequency pixels (Fig.~\ref{fig:wilson_vs_gaussian}).
We find that the frequency domain support $\sigma_f$ is increased by a factor of $\sim1.33$ with the Gaussian window.
Since the Gaussian window decays exponentially in frequency, it must be truncated.
To do so, for the frequency-domain transform we retain $q$ half-bands of the exponential tail.
We use $q = 8$ for the transforms in this work.
This $q$ is distinct from the $q$ of Ref.~\cite{Cornish2020}, which sets the half-length of the truncated time-domain Meyer kernel in units of $\Delta T$.

Figure~\ref{fig:WDM_WDG} shows the WD transformation with the Meyer and the Gaussian windows on different kind of signals.
Being more compact in time, the Gaussian window provides a better localization in time of short-duration signals, while the representations of monochromatic and slowly-chirping signals are less compact in frequency with respect to the Meyer window.

\subsection{Spectral Leakage}

It is expected that future generation detectors like LISA will have frequent data loss or unusable data that will introduce gaps~\cite{Pearson2026}.
Because edges due to gaps induce spectral leakage in the wavelet domain sensu Fourier analysis, biases could be introduced into Bayesian inference without management~\cite{Baghi2019,Pearson2026}.
A compelling solution to gaps is Bayesian data augmentation where artificial data is imputed into gaps using a distribution conditioned on the observed data~\cite{Baghi2019}.
We show that because the Gaussian window kernel is more compact in time, the extent of spectral leakage in the time direction is not as spread as the Meyer window.
This may confer advantages when managing heavily gapped or glitchy data, by reducing the sizes of matrices involved in the imputing of data in gaps.
Ref.~\cite{Pearson2026} describes the construction of the conditional distribution used to draw imputing data in gaps; the square matrices scaled to the wavelet kernel would be reduced by the same factor.
Since the kernel length can be conservatively reduced by approximately a factor of 2, we can therefore expect savings in matrix operations of about 4 or 8 (for $N^2$ and $N^3$ operations).
This reduced kernel also might prove advantageous for initializing noise estimation, since the impact of gaps becomes more localized in time. 

We define spectral leakage as the residual between the transform of complete data and gapped data, when a gap or edge boundary coincides with that of the WD time pixels.
This allows us to cleanly separate out the effect of spectral leakage and missing data (since the two cannot be separated for a boundary within a WD time pixel).
We note that since the WD wavelets satisfy Parseval's theorem~\cite{Johnson2026}, the total energy of the spectral leakage is conserved between window choices; only the time-frequency distribution of the energy is changed.

\begin{figure*}
    \centering
    \includegraphics[width=\linewidth]{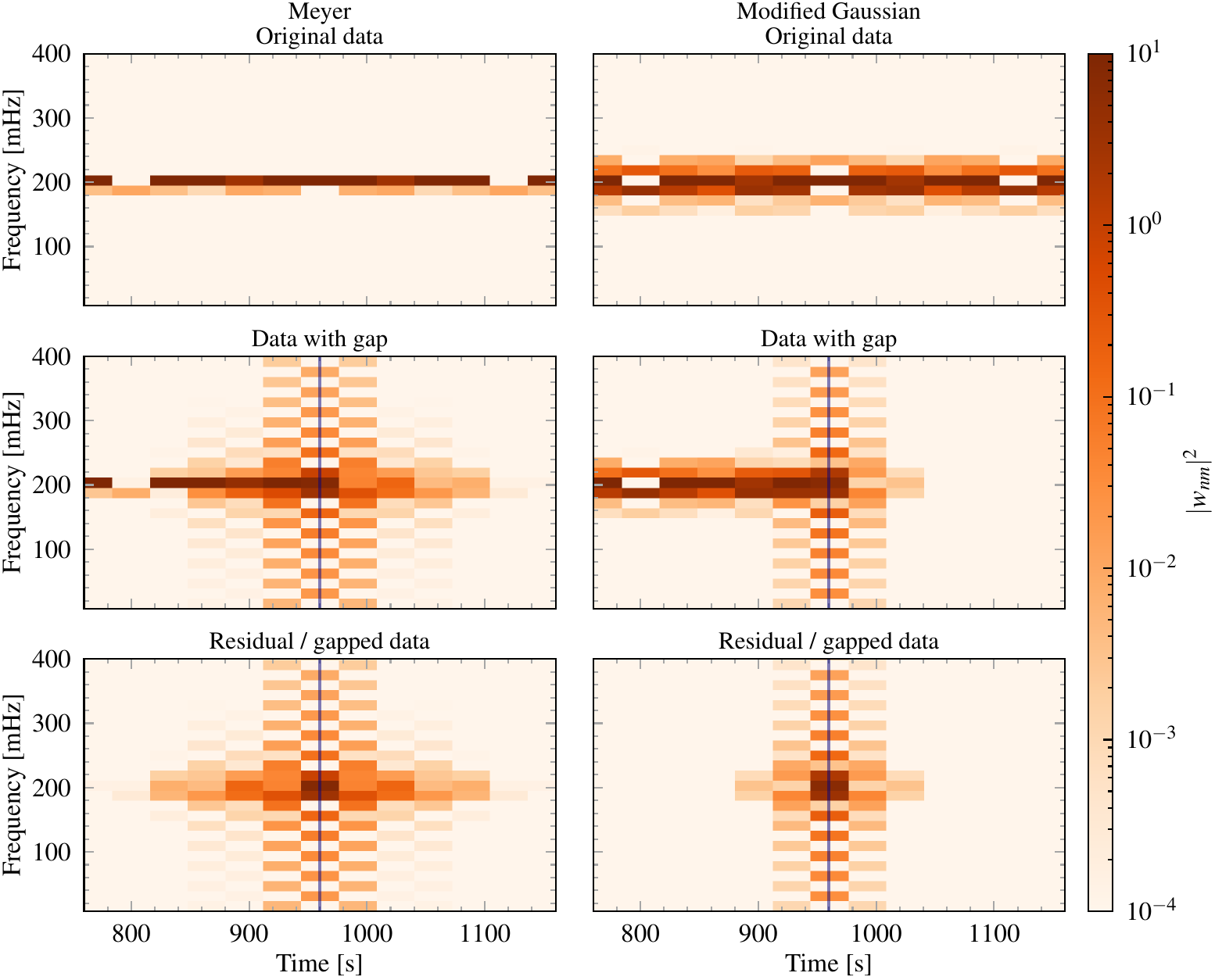}
    \caption{Effect of a data gap for the Meyer (\textit{left column}) and modified Gaussian (\textit{right column}) windows.
    The rows show the wavelet power $|w_{nm}|^2$ of a monochrome sinusoid signal at $200$~mHz for complete data, for data with an edge, and for the residual between the two transforms; we subtracted only the portion of the signal present in the gapped data (for times prior to the edge), in the top, middle, and bottom rows, respectively.
    The blue vertical line marks the edge at $960$~s, after which the data are zeroed, and the gap extends beyond the frame.
    The residual of the modified Gaussian window is confined more tightly to the edge, while the Meyer residual spreads further into the surrounding data.
}
    \label{fig:WDM_WDG_gap}
\end{figure*}

Figure~\ref{fig:WDM_WDG_gap} shows this effect for a monochrome sinusoid signal with an edge aligned with the wavelet time pixels.
We show the complete data (top row), the data with an edge (middle row), and in the bottom row, the residual of the transforms; we subtracted only the portion of the signal present in the gapped data (for times prior to the edge), for both windows.
We verify numerically that the residual energy is the same for both transforms.
A visual inspection confirms that the spectral leakage frequency distribution is comparable, while in the time direction the spectral leakage is concentrated more closely to the edge for the modified Gaussian window.
Note that the spectral leakage is symmetric and also extends into the region of removed data.

\subsection{WDG Noise-Covariance Matrix}
\label{sec:covariance}

It has been shown that the noise covariance matrix for the WDM basis is near-diagonal, with small off-diagonal terms that depend on the gradients of the dynamic spectrum power spectrum, $S(f,t)$, across the wavelet pixels~\cite{Cornish:2025awt,Cornish:2026tjt}.
The story is much the same for the WDG basis.
The leading off diagonal terms have similar magnitudes to those in the WDM basis.
For the symmetric choice $\nu=0.5$ the covariance structure is, unsurprisingly, more symmetric in the time and frequency offsets than for the WDM basis. 

The noise covariance can be written using the compact notation,
\begin{equation}
S_{\Delta m,\Delta n}\equiv S_{mn,(m+\Delta m)(n+\Delta n)}\,,
\end{equation}
\noindent where $(m,n)$ label wavelets centered at frequency and time $(f_m,t_n)$, and the offsets $(\Delta m,\Delta n)$.
The symbols,
\begin{equation}
\epsilon=\pm1,\qquad \tau=\pm1
\end{equation}
\noindent label the sign of a frequency or time offset.
The covariances can be computed by Taylor expanding the dynamic spectrum about the center of the reference WDG pixel:
\begin{equation}
S(f,t)=S_0
\sum_{p,q\ge0}
\frac{c_{pq}}{p!\,q!}\delta f^p\delta t^q,
\end{equation}
where,
\begin{equation}
S_0=S(f_m,t_n),\qquad
\delta f=\frac{f-f_m}{\Delta F},\qquad
\delta t=\frac{t-t_n}{\Delta T}.
\end{equation}
Equivalently, $c_{00}=1$ and,
\begin{equation}
c_{pq} =
\frac{\Delta F^p\Delta T^q}{S_0}
\left.
\frac{\partial^{p+q}S}
{\partial f^p\partial t^q}
\right|_{(f_m,t_n)} .
\end{equation}
The pure-frequency coefficients are $c_{p0}$, the pure-time coefficients are $c_{0q}$, and the leading mixed coefficient is $c_{11}$.
We use the normalization $S_{0,0}=S_0$ and the dimensionless ratio,
\begin{equation}
R_{\Delta m,\Delta n}\equiv \frac{S_{\Delta m,\Delta n}}{S_0}.
\end{equation}
For $\nu=0.5$ the WDG basis is almost perfectly symmetric under interchange of time and frequency.

The dimensionless noise covariance $R_{\Delta m,\Delta n}$ can be expanded numerically using a Gram matrix and the Taylor expansion of $S(f,t)$ (see App.~A of Ref.~\cite{Cornish:2026tjt} for details).
The diagonal entry is,
\begin{equation}
R_{0,0}
=1+0.08146\left(c_{20}+c_{02}\right)
+0.00363\left(c_{40}+c_{04}\right)+\cdots .
\end{equation}
The coefficient $0.08146$ is one half of the squared rms width of the Gaussian window, Eq.~\eqref{eq:variances}, in units of either $\Delta F$ or $\Delta T$.
For the nearest time neighbor in the same frequency layer,
\begin{align}
R_{0,\tau}
&=(-1)^{n+1}
\left[
\tau\left(0.24047\,c_{10}+0.01004\,c_{30}\right)\right. \nonumber \\
& \left. +0.12023\,c_{11}
+\cdots
\right].
\end{align}
For the nearest frequency neighbor in the same time row,
\begin{align}
R_{\epsilon,0}
&=(-1)^{n+m}
\left[
\epsilon\left(0.24047\,c_{01}+0.01004\,c_{03}\right)\right. \nonumber \\
& \left. +0.12023\,c_{11}
+\cdots
\right].
\end{align}
The first term in $R_{0,\tau}$ is driven by the local frequency slope, while the first term in $R_{\epsilon,0}$ is driven by the local time slope.
The mixed derivative $c_{11}$ contributes to both axial offsets with the same magnitude.
Note that while the leading coefficients in the expansion are relatively large, setting the gradients to unity, $c_{01}=1$ or $c_{10}=1$, correspond to 100\% changes in the noise level across a single time-frequency pixel, which is rather extreme.

For the four nearest diagonal pixels,
\begin{align}
R_{\epsilon,\tau}
&=(-1)^m
\left[
-\epsilon\left(0.10456\,c_{10}+0.01940\,c_{30}\right)\right. \nonumber \\
& \left. +\tau\left(0.10456\,c_{01}+0.01940\,c_{03}\right)
\right. \nonumber\\
&\left.
-0.05228\,c_{20}
+0.05228\,c_{02} \right. \nonumber \\
&\left.
-0.00534\,c_{40}
+0.00534\,c_{04}
+\cdots
\right],
\end{align}
while the leading $c_{11}$ contribution to $R_{\epsilon,\tau}$ is zero at the displayed precision.
These terms show the expected antisymmetry between the frequency-gradient and time-gradient pieces.

As the offsets get larger the correlations get smaller.
For a separation of two time rows in the same frequency layer,
\begin{align}
R_{0,2\tau}
&=
-0.02173\,c_{20}
-0.02165\,c_{02}
-\tau\,0.02165\,c_{03}
\nonumber\\
&\hspace{0.45in}
+0.0000815\,c_{40}
-0.01443\,c_{04}
+\cdots .
\end{align}
For a separation of two frequency layers in the same time row,
\begin{align}
R_{2\epsilon,0}
&=
-0.02165\,c_{20}
-0.02173\,c_{02}
-\epsilon\,0.02165\,c_{30}
\nonumber\\
&\hspace{0.45in}
-0.01443\,c_{40}
+0.0000815\,c_{04}
+\cdots .
\end{align}
The small difference between $0.02173$ and $0.02165$ is a finite discrete-window effect; in the continuum symmetric limit these are the same coefficient.


\begin{figure*}
    \centering
    \includegraphics[width=0.75\linewidth]{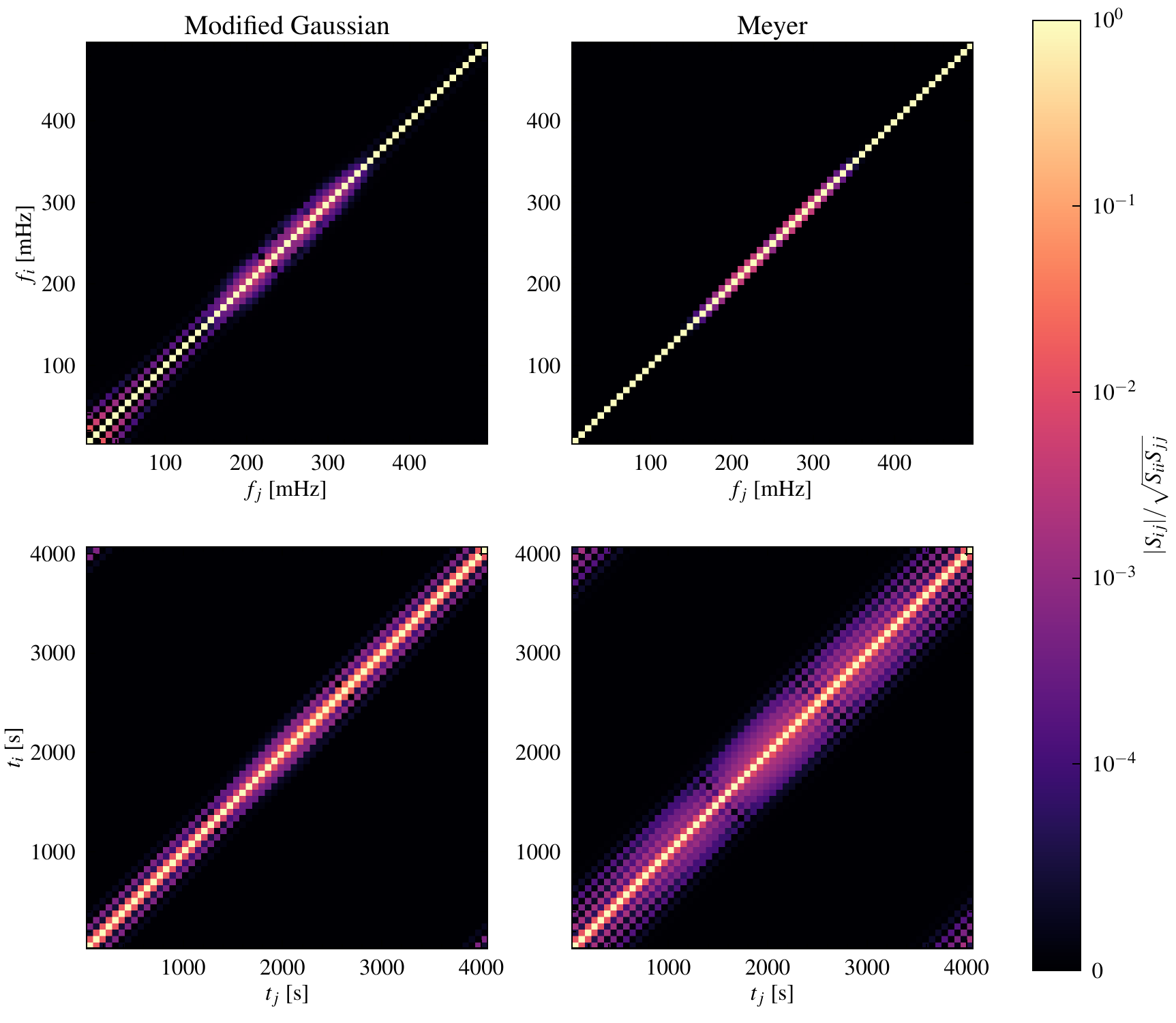}
    \caption{The normalized noise covariance matrix for the modified Gaussian (\textit{left}) and Meyer (\textit{right}) windows, shown for constant time (\textit{top row}) and frequency (\textit{bottom row}) slices.
    Note that in the constant time slice, the Meyer window shows more local correlations in frequency with smaller off-diagonal elements; the opposite is true in the constant frequency slice.}
    \label{fig:noise_covariance}
\end{figure*}

In Fig.~\ref{fig:noise_covariance} we show the normalized noise covariance matrices for both the Meyer and modified Gaussian windows.
For this visual comparison we use a Gaussian bump as the dynamic spectrum, as in Ref.~\cite{Cornish:2026tjt}.
We take constant time and frequency slices to see the influence of window compactness on correlations in both dimensions.
For a constant time slice, we see that the modified Gaussian predictably produces more (albeit modest) off-diagonal terms than the Meyer window, and conversely in the constant frequency slice.

\begin{figure*}
    \centering
    \includegraphics[width=\linewidth]{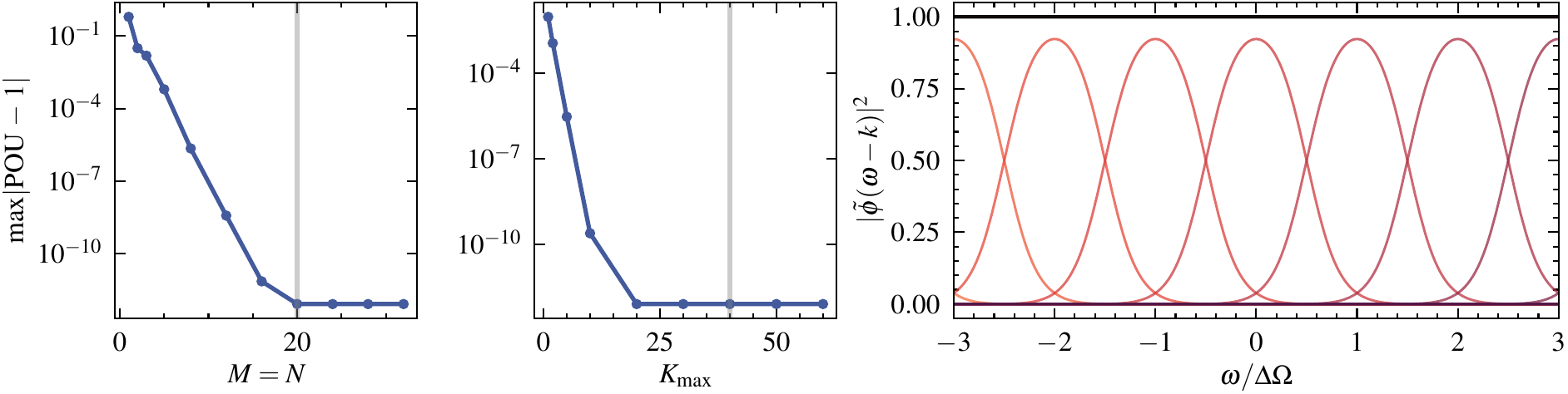}
    \caption{(\textit{left}) Deviation from partition of unity (POU) varying the number of $M$ and $N$ terms in the summation of Eq.~\eqref{eq:gaussian window}.
    The maximum is computed over the window support.
    (\textit{center}) Deviation from partition of unity (POU) varying the number $K_{\mathrm{max}}$ of terms in the summation of Eq.~\eqref{eq: coeffs 1}.
    The vertical gray lines indicate the parameter choice.
    (\textit{right}) Numerical evaluation of partition of unity for $M = N = 20$ and $K_{\mathrm{max}} = 40$.
    The colored lines represent the squared modified Gaussian windows, $|\tilde{\phi}(\omega - k\,\Delta\Omega)|^2$ for integer $k$.
    The black line shows the sum of these squared windows and is $\simeq$ 1, as expected.
    For $\nu=0.5$ the modified Gaussian window is the same form in time and frequency so POU is verified in the time domain as well.}
    \label{fig:POU}
\end{figure*}

\section{Conclusion}

In this work, we have investigated the Gaussian window as a WD-compatible construction that is more compact in time-frequency than the standard Meyer window choice.
We show this using several measures, including the time-frequency area (evaluated analytically and numerically), and the point spread function (App.~\ref{appx: PSF}).
Notably, the Gaussian window comes within $\sim2.4\%$ of saturating the fundamental Heisenberg-Gabor time-frequency area limit; based on this result and motivated by the form of the analytic time-frequency area Eq.~\eqref{eq:sigmaf_analytic}, we conjecture that the Gaussian window achieves the minimum time-frequency area of any WD wavelet window.

The analytic result relies on two properties of the construction.
First, every moment of the window is a Gaussian integral, which gives the frequency variance in closed form as the variance of the Gaussian kernel plus a lattice correction from the orthogonalization (Eq.~\eqref{eq:sigmaf_analytic}).
Second, the window at shape parameter $\nu$ in time is a rescaled copy of the window at $1/(4\nu)$ in frequency (Eq.~\eqref{eq:nu_duality}), which fixes the time variance and makes the product stationary at the symmetric choice $\nu = 1/2$, where our scan finds its minimum.
There the two widths are equal, $\sigma_t = 0.40\,\Delta T$ and $\sigma_f = 0.40\,\Delta F$, and the variance product is $1.0236/(4\pi)$. 
No Meyer window approaches this value, since its product is $0.107$ at the standard choice, $34\%$ above the bound, and never below $13\%$ over the steepness parameters and flat-top fractions we examined.

We further show that the Gaussian window is a useful expansion of the toolkit available to \gw researchers, especially in preparation for the LISA mission.
This window's time-frequency compactness may confer advantages over the standard Meyer window when analyzing merging \gw signals, and managing glitch- or gap-dense data.
This is achieved through savings in the time-domain wavelet transform kernel of about a factor of 2.5-3.
This shorter kernel also reduces the computational cost necessary for Bayesian data augmentation.

The noise covariance matrix of the WDG basis remains near-diagonal, as for the WDM basis, with off-diagonal terms of similar size that are nearly symmetric in time and frequency for $\nu = 1/2$ (Sec.~\ref{sec:covariance}), so the treatment of non-stationary noise developed for the WDM basis carries over.
The shape parameter can be adjusted away from $\nu = 1/2$ when an analysis favors one domain, toward smaller $\nu$ for gap-dense data and toward larger $\nu$ for long-lived, nearly monochromatic signals, at the cost set by Eq.~\eqref{eq:sigma_swap}.

Several questions remain open.
The conjecture that no WD window is more compact than the Gaussian window is supported by the structure of Eq.~\eqref{eq:sigmaf_analytic} but not proven, and a proof, or a counterexample, would resolve the open question of the optimal window~\cite{Cornish:2025awt}.
Independently of the conjecture, any improvement in this product by another WD window is limited to $2.4\%$, the excess over the bound for the Gaussian window.
We leave a detailed study of the cost of Bayesian data augmentation with the shorter WDG kernel, and of the use of the window in low-latency searches with ground-based detectors, to future work.

\begin{acknowledgments}
We would like to thank the organizers of the 4th LISA Sprint held in May 2026 in Bozeman, Montana where this work was initiated. 
We further extend our gratitude to Katerina Chatziioannou and Aaron Johnson for their valuable conversations and contributions to this work during the LISA Sprint.
The agentic coding tools Codex and Claude were used to refine and produce much of the WDG transform code.
We have reviewed and tested the code and take responsibility for its accuracy.
N.~Pearson and N.~J.~Cornish were supported by NASA LISA Preparatory Science Grant 80NSSC19K0320 and Simons Foundation award SFI-MPS-BH-00012593-04. 
S.~Bini was supported by NSF Grant PHY-2309200.
M.~Çalışkan was supported by the Rowland Research Fellowship of Johns Hopkins University.
\end{acknowledgments}

\appendix

\section{Verifying Partition of Unity and Orthogonality}
\label{appx: POU and orthog}

\begin{figure}
    \centering
    \includegraphics[width=\linewidth]{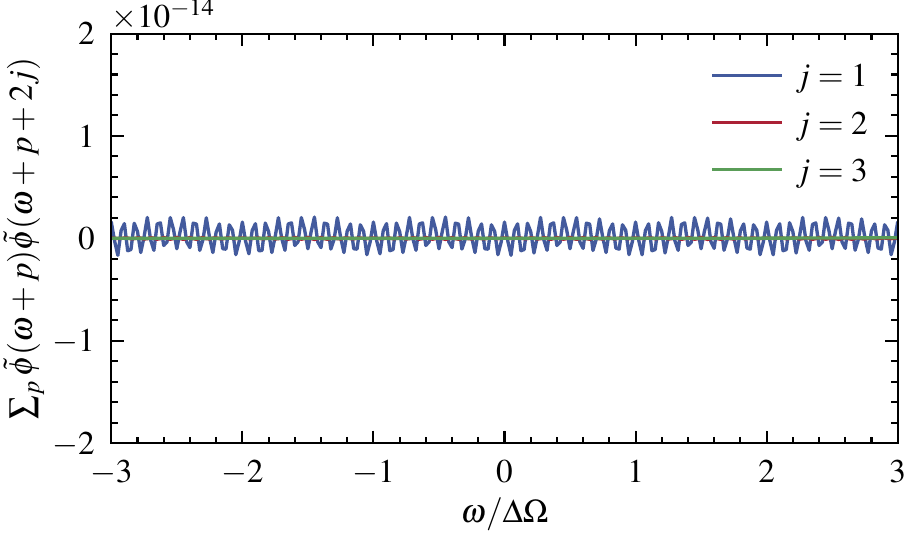}
    \caption{The three solid curves show the left side of Eq.~\eqref{eq:DJJcondition} as a function of the angular frequency $\omega$ in units of the pixel width $\Delta\Omega$, for $j=1$ (blue), $j=2$ (red), and $j=3$ (green), computed with $M=N=20$, $K_{\mathrm{max}}=40$, and the sum over $p\in[-16,16]$.
    The residuals remain below $3\times10^{-15}$, at the level of double floating-point precision, confirming orthogonality is satisfied numerically.}
    \label{fig:orthogonality}
\end{figure}

In order to have an orthogonal WD transform, we require that the window $\tilde{\phi}$ satisfies~\cite{Daubechies1991,Johnson2026},

\begin{equation}
\sum_{p \in \mathbb{Z}} \tilde{\phi}(f+p \Delta F) \tilde{\phi}(f+(p+2 j) \Delta F)=\frac{\delta_{j0}}{\Delta F}\,, \quad \forall j \in \mathbb{Z}\,.
\label{eq:DJJcondition}
\end{equation}

\noindent The case where $j=0$ is the statement of partition of unity, and $j\neq0$ states the cancellation of cross-band windows.
We require both partition of unity and orthogonality in order to have a uniform distribution in power and a lossless transform, respectively. 
The Gaussian window satisfies both conditions by construction~\cite{Daubechies1991}, but within an error determined by the number of terms included in the sums.

We verify the modified Gaussian satisfies partition of unity numerically, by evaluating Eq.~\eqref{eq:DJJcondition} for $j=0$.
We compute the deviation from partition of unity varying the number $M,\,N$ and $K_{\mathrm{max}}$ of terms in Eq.~\eqref{eq:gaussian window} and Eq.~\eqref{eq: coeffs 1}.
The left and center panels in Fig.~\ref{fig:POU} show that by using $M=N=20$ and $K_{\mathrm{max}}=40$, the deviation from partition of unity becomes negligible ($<10^{-11}$).
Using these values, we visualize the partition of unity in the right panel in Fig.~\ref{fig:POU}: the sum of the squared windows is $\simeq 1$. 

In Fig.~\ref{fig:orthogonality}, we demonstrate numerically that the modified Gaussian window satisfies the orthogonality requirement, computing Eq.~\eqref{eq:DJJcondition} for $j\neq 0$. 
We generate modified Gaussian windows using $M=N=20$ and $K_{\mathrm{max}}=40$, and the sum over $p\in[-16,16]$. 

\section{Point Spread Function}
\label{appx: PSF}

\begin{figure}
    \centering
    \includegraphics[width=\linewidth]{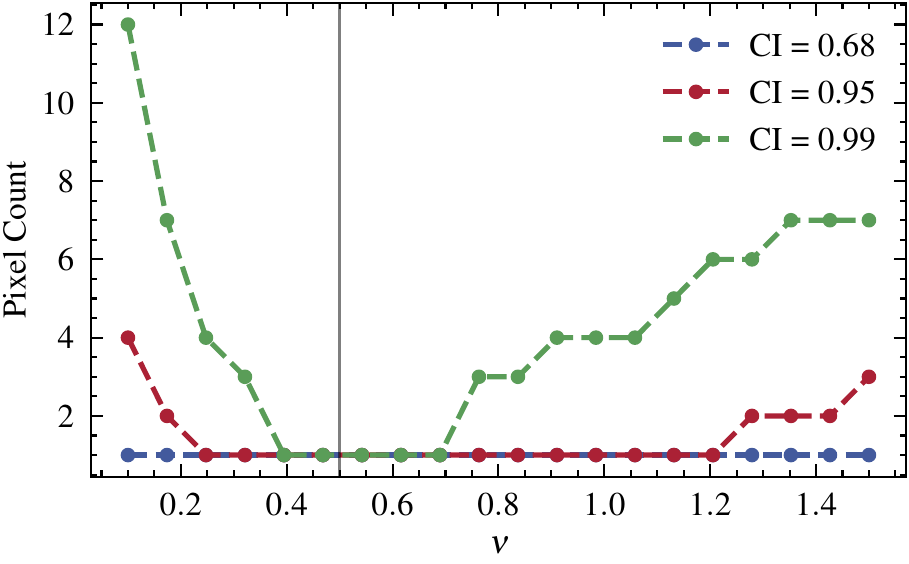}
    \caption{The modified Gaussian window point spread function (PSF) for a symmetric Morlet-Gabor wavelet ``point source'' injection.
    Plotted are the number of pixels that contain 68\%, 95\%, and 99\% of the injected energy for various parameter $\nu$ choices.
    The vertical bar indicates the standard parameter choice $\nu = 0.5$.}
    \label{fig:gaussian_PSF}
\end{figure}
\begin{figure}
    \centering
    \includegraphics[width=\linewidth]{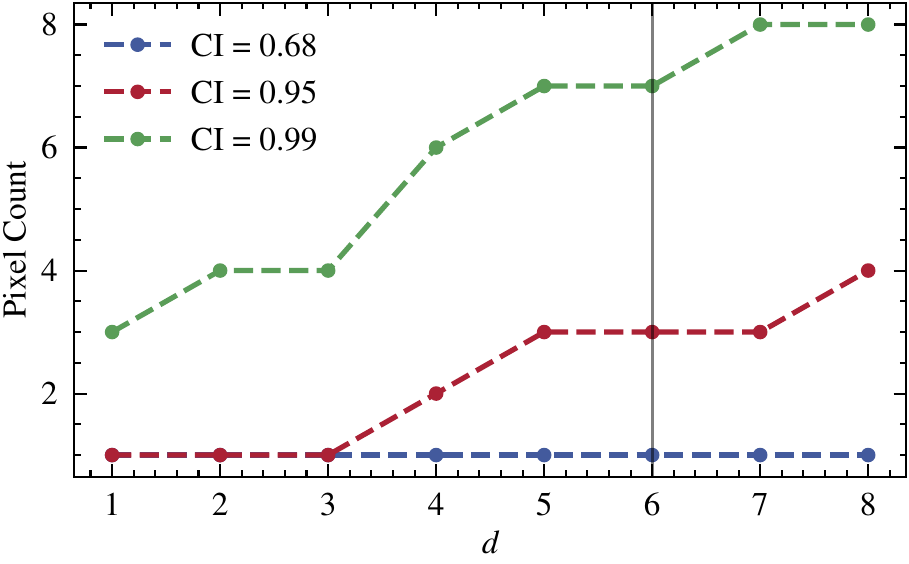}
    \caption{Meyer window point spread function for the symmetric Morlet-Gabor point source of Fig.~\ref{fig:gaussian_PSF}, as a function of the steepness parameter $d$ at $A=0$.
    The dashed curves with circular markers show the number of pixels containing $68\%$ (blue), $95\%$ (red), and $99\%$ (green) of the injected energy, and the vertical gray bar marks the standard choice $d=6$, where the $99\%$ count is seven pixels.}
    \label{fig:meyer_PSF1}
\end{figure}

\begin{figure}
    \centering
    \includegraphics[width=\linewidth]{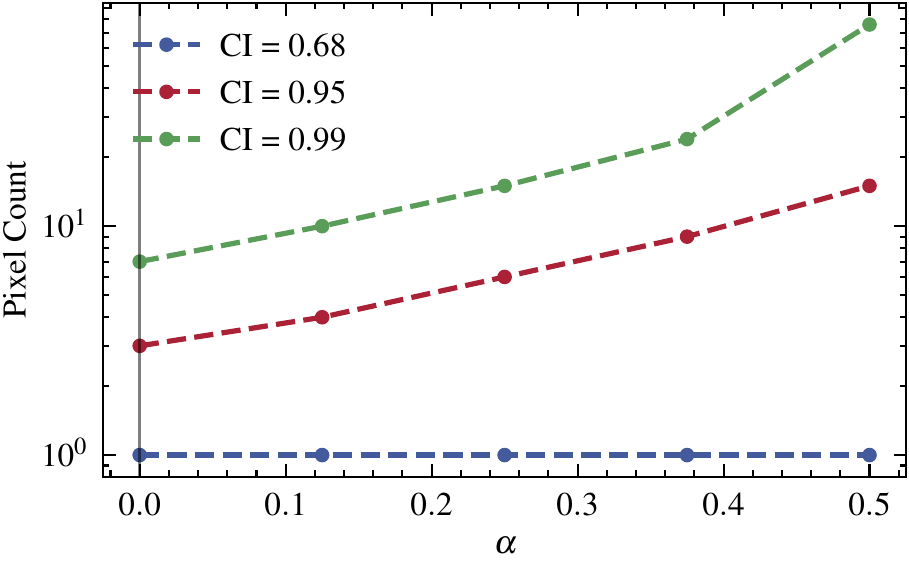}
    \caption{Same as Fig.~\ref{fig:meyer_PSF1}, but varying the flat-top fraction $\alpha = A/\Delta\Omega$ at $d=6$ under the constraint $2A+B=\Delta\Omega$, with the pixel count on a logarithmic scale.
    The vertical gray bar marks the standard choice $\alpha=0$.
    The $95\%$ and $99\%$ counts grow by an order of magnitude as $\alpha \rightarrow 1/2$, where the roll-off width vanishes and the window approaches a sharp box.}
    \label{fig:meyer_PSF2}
\end{figure}

To further illustrate the difference in time-frequency compactness between the Meyer and Gaussian windows, we estimate the Point Spread Function (PSF) as in optics.
We use a Morlet-Gabor wavelet~\cite{Cornish2014} injection as the idealized time-frequency ``point source'', since it has the minimum possible variance product.
The wavelet has been constructed to have the same spread relative to the WD wavelet pixel widths, $\Delta T,\,\Delta F$ (i.e. time domain variance of the wavelet $\tau^2 = \Delta T/(\pi \Delta F)$).
With this lens, we see that the Gaussian window very effectively concentrates energy in time-frequency, capable of containing 99\% of the injected energy in one pixel, cf. Fig.~\ref{fig:gaussian_PSF}.
We note that this measure of compactness qualitatively mirrors that of the time-frequency variance product, Fig.~\ref{fig:area}.
The Meyer window PSFs by comparison show that it spreads 99\% of the energy across seven pixels for the standard parameter choices.
Figures~\ref{fig:meyer_PSF1} and \ref{fig:meyer_PSF2} show the Meyer pixel counts as functions of the steepness parameter $d$ and of the flat-top fraction $\alpha = A/\Delta\Omega$, respectively.

\bibliographystyle{utphys}
\bibliography{refs}

\end{document}